\documentclass[11pt]{article}
\usepackage{epsf,amssymb}
\newcommand{\ZZ}{\hbox{{\sf Z{\hbox to 2pt{\hss\sf Z}}}}}
\newcommand{\PP}{\hbox{{\sf I{\hbox to 6pt{\hss\sf P}}}}}
\newcommand{\bfig}{\begin{figure}}
\newcommand{\efig}{\end{figure}}

\newcommand{\One}{\hbox{{\rm 1{\hbox to 1.5pt{\hss\rm1}}}}}
\newcommand{\virg}{\; \; ,}

\newcommand{\pu}{\; \; .}
\newcommand{\nn}{\nonumber}
\newcommand{\ri}{\right}
\newcommand{\lf}{\left}

\newcommand{\Y}{\Upsilon}
\newcommand{\Z}{{\cal Z}}

\newcommand{\K}{{\cal K}}

\newcommand{\eq}{\begin{equation}}
\newcommand{\en}{\end{equation}}
\newcommand{\bea}{\begin{eqnarray}}
\newcommand{\eea}{\end{eqnarray}}

\newcommand{\ba}{\begin{array}}
\newcommand{\ea}{\end{array}}

\newcommand{\CC}{{\hbox{\rm C\kern-0.5em{$\sf I$}}}}
\newcommand{\II}{\hbox{{\rm l{\hbox to 1.5pt{\hss\rm l}}}}}
\newcommand{\RR}{{\hbox{$\rm\textstyle I\kern-0.2em R$}}}

\newcommand{\NP}[1]{Nucl.\ Phys.\ {\bf #1}}
\newcommand{\PL}[1]{Phys.\ Lett.\ {\bf #1}}

\newcommand{\IJMP}[1]{Int.\ J.\ Mod.\ Phys.\ {\bf #1}}

\newcommand{\ABZ}{A.B.Zamolodchikov}
\newcommand{\AlBZ}{Al.B.Zamolodchikov}

\newcommand{\fract}[2]{{\textstyle\frac{#1}{#2}}}

\newcommand{\opnup}[1]{\renewcommand{\\}{\\[50 pt]}}
\renewcommand{\bar}{\overline}
\renewcommand{\tilde}{\widetilde}

\renewcommand\hat{\widehat}
\newcommand\abstracts[1]{
\begin{center}
{\begin{minipage}{4.2truein}
                 \footnotesize
                 \parindent=0pt #1\par
                 \end{minipage}}\end{center}
                 \vskip 2em \par}
\begin{document}
\begin{flushright}
DFTT- 5/99 \\
ITFA 99-3 \\
{\tt hep-th/9902094} \\
February 1999
\end{flushright}
\begin{center}
{\bf A TOPOLOGICAL INVARIANT OF RG FLOWS}
\end{center}
\begin{center}
{\bf IN 2D INTEGRABLE QUANTUM FIELD THEORIES}
\end{center}
\vskip .1 cm
\begin{center}
{\sl R.Caracciolo$~^\dagger$,  F. Gliozzi$~^\dagger$
and R.Tateo$~^\ddagger$}
\end{center}
\begin{center}
{\small $\dagger~$Dip. di Fisica Teorica, Universit\`a
di Torino, v. P.Giuria 1, 10125 Torino, Italy}\\
{\small $\ddagger~$Univ. van Amsterdam, Inst. voor
Theor. Fys., 1018 XE Amsterdam, The Netherlands}
\end{center}
\vskip .2 cm
\abstracts{ We construct a topological invariant of the
renormalization group
trajectories of a large class of 2D quantum integrable
models, described by the thermodynamic Bethe ansatz
approach.
A geometrical description of this invariant in terms of
triangulations of three-dimensional manifolds is
proposed and associated dilogarithm identities are
proven.}
\section{Introduction}
Many quantum integrable  2D models can be considered
as  perturbations of a Conformal Field Theory.
The RG evolution of these models is described, in the
thermodynamic Bethe ansatz (TBA) approach~\cite{Zam1}
by the ground state energy $E(R)$ of the system in a
infinite cylinder of radius $R$, through an  equation of the form
$ E(R)=\frac {-1}{2\pi}\sum_{a=1}^{N}\int d\theta\,
\nu_a(\theta)\log\left(1+Y_a(\theta)\right)$,
where $\theta$ is the rapidity, $a$ labels the
pseudoenergies~(cf.~\cite{KM}), $Y_a(\theta)$ are
$R$-dependent functions, determined by a set
of coupled integral equations known as TBA equations;
 the $\nu_a$'s are elementary functions of $R$ and
 $\theta$, related to  the asymptotic
 behaviour of the $Y_a(\theta)$'s for
 $\theta\to\pm\infty$.
An important observation of \AlBZ \cite{Zf} was that any solution
$\{ Y_a(\theta)\}$ of the TBA equations satisfies a
set of simple functional equations, called the Y-system.
Conversely, it can be shown that under appropriate
analytic and asymptotic  restrictions the Y-system is
equivalent to the set TBA equations.  For instance, the
Y-system describing the flow from the tricritical Ising
model to the critical Ising one is given by
\eq
Y_1(\theta+\frac{i\pi}2)Y_1(\theta-\frac{i\pi}2)=
1+Y_2(\theta)~,~~Y_2(\theta+\frac{i\pi}2)Y_2(\theta-\frac{i\pi}2)=
1+Y_1(\theta)~~.
\label{A2}
\en
This system, like any other Y-system, can be considered
as a  recursion relation in the
imaginary rapidity. To simplify the notations we can
put $\Upsilon_n(R,\theta)=Y_{\frac{3}2+\frac{(-1)^n}{2}}(\theta+n \frac{i\pi}2)$, then
the eq.~(\ref {A2}) becomes
\eq \Upsilon_{n-1}\Upsilon_{n+1}=
1+\Upsilon_n ~.
\label{penta}
\en
It is immediate to verify that this  Y-system is
{\sl periodic}: $ \Upsilon_{n+5}=
\Upsilon_n$, i.e.
$ Y_1(\theta+\frac 52 i\pi)=Y_2(\theta)~,
~~Y_2(\theta+\frac 52 i\pi)=Y_1(\theta)~~$.
In ref. \cite{Zf} it was also conjectured that a similar
periodicity $Y_a(\theta+Pi\pi)=Y_{\bar a}(\theta)$
where the period $P$ is related to the dimension
of the perturbing operator by $\Delta=1-1/P$.
In the UV limit $R\to 0$ the vacuum energy $E(R)$ is expected
to behave as $E(R)\sim -\pi \tilde c/6R$, where
$\tilde c$ is the effective central charge of th
UV fixed point. In general one finds
\eq
\tilde c=\frac 6{\pi^2}\sum_{a=1}^N\left[
L\left(\frac{Y_a(-\infty)}{1+Y_a(-\infty)}\right)
-L\left(\frac{Y_a(\infty)}{1+Y_a(\infty)}\right)
\right]~,
\label{cuv}
\en
where $L(x)$ is the Rogers dilogarithm. It can be
defined
 as the unique function three times differentiable
which satisfies the following five term relationship,
known also as the Abel functional equation (see for instance
\cite{Kirillov})
\eq
L(x)+L(y)+L\left(\frac{1-x}{1-xy}\right)+L(1-xy)+
L\left(\frac{1-y}{1-xy}\right)=3\,L(1)
\virg
\label{SAK}
\en
with $0\leq x,y\leq1$ and the normalization $L(1)=
\frac{\pi^2}6$.
This equation has a $\ZZ_5$ symmetry \cite{Zagier1,GT1}
which becomes manifest rewriting it in the form \cite{GT1}
\eq \sum_{n=1}^5L\left(\frac{\Upsilon_n}{1+\Upsilon_n}
\right)=3\,L(1)~,
\label{SAK1}
\en
where the $\Upsilon_n$'s fulfil the recursion
relation (\ref {penta}).
Clearly eq.~(\ref{SAK1}) defines  an invariant
of the RG flow of the tricritical Ising system,
being the $L$ arguments non-trivial functions of $R$ and $\theta$,
while the r.h.s. is  constant along the RG trajectory.
This behaviour suggests
\cite{GT1} the existence of  large families
of  functional equations, which are generalizations
of (\ref{SAK1}).
The simplest examples are classified in terms of an ordered  pair
$G \times H$ of $ADET$ ($T_n=A_{2n}/\ZZ_2$)  Dynkin
diagrams \cite{RTQ}
\eq
\Y_a^b(n-1)\Y_a^b(n+1)=
\prod_{c=1}^{r_G} (1+\Y_c^b(n))^{G_{ac}}
\prod_{d=1}^{r_H} \lf(1+{1 \over \Y_a^d(n)} \ri)^{-H_{bd}}
\pu
\label{YS}
\en
Here $G_{ac}$ and $H_{bd}$ are the adjacency matrices of the appropriate
Dynkin diagrams whereas $r_G$ and $r_H$ are the ranks of the corresponding
algebras. The solutions to this system are
conjectured to be periodic functions:
$ \Y_a^b(n+P)=\Y_{\bar{a}}^{\bar{b}}(n)\,,~ P=h+g\,,$
with $\bar{a}$ and $\bar{b}$ the conjugate nodes to $a$ and $b$,
while  $h$ and $g$ are the dual Coxeter numbers.
Then, the $\Y$'s
solving eq.~(\ref{YS}) are, according to our conjecture,
arguments of
functional equations for the Rogers dilogarithm, $L(x)$, of the following form
\eq
\sum_{a=1}^{r_G} \sum_{b=1}^{r_H} \sum_{n=0}^{h+g-1}
L\lf( {\Y_a^b(n) \over 1+\Y_a^b(n)} \ri)=r_G r_H g L(1)
\pu
\label{adedilog}
\en
Notice that since $L(x)$ is a multi-valued function,
  the r.h.s of eq.~(\ref{adedilog})
cannot be fixed unambiguously, unless a consistent
choice of the sheet of the associated Riemann surface
is implemented. The quoted value corresponds to
solutions of the recursion relations with
all the $\Y$'s  $>0$.
It is interesting that the  $A_N \times A_1$  ($N=1,2,3$) identities
reduce
respectively
 to the Euler, Abel and Newman functional equations
(cf.~\cite{Lewin1} ) when symmetric conditions on the conjugated nodes are
chosen.

 Other examples of Y-systems with
their corresponding dilogarithm identities were found
and discussed in ref. \cite{RT1}.
Furthermore, two new families  of Y-systems and dilogarithm functional
equations, classified in terms of a pair of co-prime integer $(p,q)$
associated
to the sine-Gordon model at rational points and to its
reductions,
were proposed and partially studied \cite{RT1,GT2}
(see \cite{DTT} for even more exotic relations).
Very soon after their discovery, the  general
solution of the $A_N \times A_1$  systems
was  found  and the corresponding
dilogarithm identities independently proven in
\cite{GT2,FS}.
We report here the solution proposed in ref. \cite{GT2},
which  was related to the  fact, already known
to Lobachevskij,
that dilogarithms define volumes of particular
tetrahedra in hyperbolic space and other
manifolds.
The seed of the idea was introduced
in 1992 in the field of integrable models
by ref.~\cite{NRT}.
Other ideas  were then expound and elaborated
in ref.~ \cite{DS}, where a manifold on which
Rogers dilogarithms define volumes of ideal
tetrahedra was identified.
\section{The geometrical picture}
The starting point of ref. \cite{GT2} was the assumption
that the Abel
equation  mimics in some way a volume calculation
of a compact manifold.
Although the existence and the nature of such a manifold
could not be further  argued, this geometrical
point of view
was rather powerful and allowed  to prove
in a rigorous way, just using eq.~(\ref{SAK}) as the only
ingredient, the $A_N \times A_1$ identities and to
solve the $Y$ system of a general class of thermally
perturbed minimal models.
Let us mention  that  after the main results of
ref.s~\cite{GT1,GT2,FS} were already appeared,
we became aware
\footnote{
We are grateful to the  referee of an earlier version
of  \cite{GT2} for a copy of Zagier's notes.}
of an  undated manuscript  by  D.Zagier 
where this geometric approach is clarified and
simplified.
We can now rephrase our construction as follows.
Consider an arbitrary triangulation of an {\sl arbitrary}
three-dimensional manifold, made by $M$ oriented
tetrahedra $\{T_i\}$.  The fact that this set forms
a triangulation means that the boundary of the
associated 3-chain is zero:
\eq
\partial\; (\sum_{i=1}^{M} T_i )=0
\virg
\en
where the boundary of the tetrahedron $T\equiv [ABCD]$ is
defined in the standard way:
\eq
\partial\; [ABCD]=[ABC]-[BCD]+[CDA]-[DAB]
\pu
\label{bound}
\en
Now associate to each vertex $i$ $(i=1,\dots N)$ a
real number \footnote{More generally, the  $x_i$'s
may be
chosen as points of an arbitrary circle of the
complex plane in order to have real cross-ratios
in eq.~(\ref{cr}).} $x_i$.
Through this map we can assign to each tetrahedron
$T$,  the cross-ratio $t=(abcd)$ of the
values associated to its vertices
\eq
T\equiv [ABCD]\to t=(abcd)~,~~~
(abcd)=\frac{(a-c)(b-d)}{(a-d)(b-c)}
\pu
\label{cr}
\en
In the following we shall use some elementary properties
of the cross-ratios:
\begin{eqnarray}
(abcd)&=&(cdab)~=~(badc)~=~1-(acbd)~=~(bacd)^{-1}  \nn \\
 (aabc)&=&1~~,~~(abbc)~=~\infty~~,~~(abcb)~=~0
\pu
\end{eqnarray}
We shall prove that
\eq
\sum_{j=1}^ML(t_j)=n\,L(1)
\virg
\label{triaid}
\en
where the integer $n$ is unambiguously fixed when the
sequence of real numbers $\{x_i\}$ is ordered in such a
way that all the $t_j$'s belong to the
interval $[0,1]$. Since the number $M$ of tetrahedra
is less or equal to the number $N$ of vertices, there
are simple algebraic relations among the
$t_j$'s. For suitable triangulations they coincide
with the Y-system of known integrable models.
Before giving a general proof of eq.~(\ref{triaid})
let us see a couple of simple instances.
Consider a  polytope made of tetrahedra in four
dimensional Euclidean space.
Well known examples of this kind
are  the three regular polytopes
5-cell (or 4-simplex) , 16-cell (dual to hypercube)
and 600-cell. They are all triangulations of $S_3$
(see for example \cite{Coxeter}).
The 5-cell of fig. 1a is formed by  the
following five tetrahedra
\[
T_1=[BACD],~T_2=[CBDE],~T_3=[DCEA],~  T_4=[EDAB],~
T_5=[AEBC]
\pu
\]
{\begin{figure}[ht]
\[\begin{array}{ll}
\epsfxsize=.37\linewidth\epsfbox{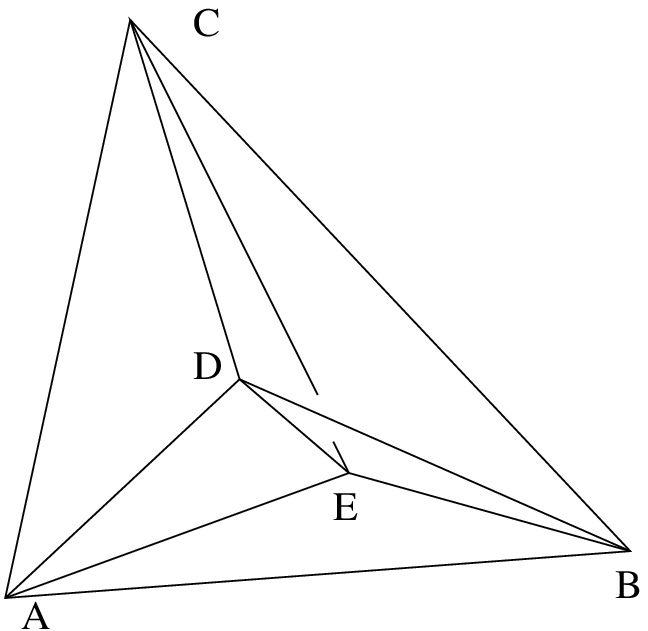}
{}&
\epsfxsize=.37\linewidth\epsfbox{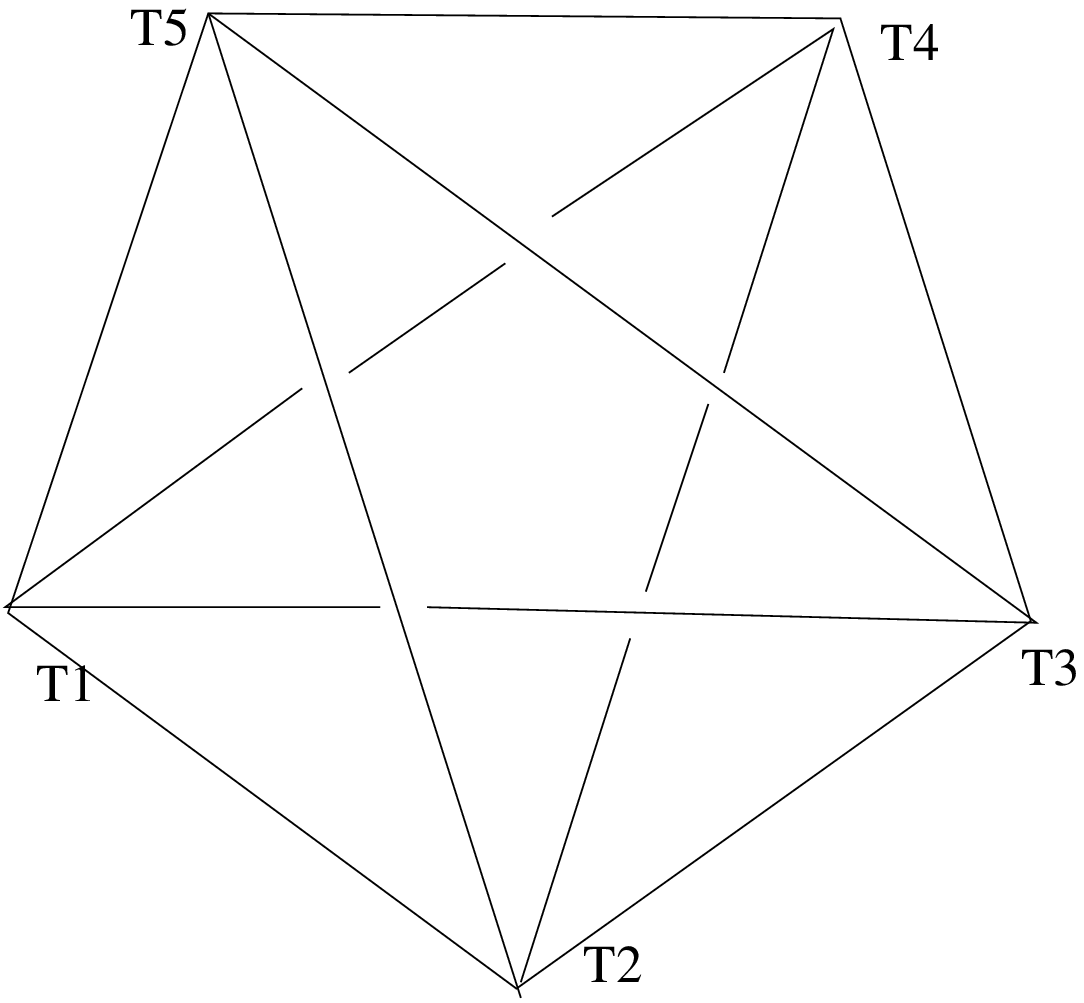}
\\
\parbox[t]{.47\linewidth}{\small 1a) $5~ cell$.}
{}~&~
\parbox[t]{.47\linewidth}{\small 1b) $dual ~polytope$.}
\end{array}\]
\label{cel}
\end{figure}}
%
The pentagonal relation~(\ref{SAK})
is then secured assigning five real numbers
$\{a<b<c<d<e\}$ to the vertex set $\{A,B,C,D,E \}$;
one immediately  end-up with  \footnote{Because of the projective
invariance of the cross-ratio, three of these values
can be fixed arbitrarily, then these five cross-ratios
depend actually only on two free parameters, as required by
eq.~(\ref{SAK}).}
\eq
L((bacd))+L((cbde))+L((dcea))+L((edab))+L((aebc))=
3 L(1)
\pu
\label{abel1}
\en
 It is instructive to draw the {\sl adjacency
 graph} of this set of tetrahedra; associating  each
 tetrahedron to a vertex an connecting with a line two
 tetrahedra sharing the same face, we end up with the
 graph of the dual polytope, which in this case is
 again a 5-cell (see fig. 1b). For each loop
 of the adjacency graph one can write an algebraic
 relation among the cross-ratios of the tetrahedra touched
 by the loop.
In particular, it is easy to verify that the $L$
arguments fulfil the recursion relation (\ref {penta}).
Moreover the well known identity $L(x)+L(1-x)=
L(1)$, known as Euler equation, can be written as
\eq
L((abcd))+L((acbd))= L(1)
\virg
\label{euler1}
\en
hence
\eq
L((bcad))+L((cdbe))+L((deca))+L((eadb))+L((abec))=
2\, L(1)
\label{abel2}
\pu
\en
In the following we shall use the short-hand notation
$L\{abcde\}=3L(1)$ for the identity (\ref{abel1})
and $\tilde L\{abcde\}=2L(1)$ for the
identity (\ref{abel2}).
As  a more engaging example, let us  consider  the
16-cell. This polytope has 8 vertices, characterized
by the  property that
each vertex is linked to all the
others but one (see fig. 2a). Ordering the vertices
in such a way that the unlinked
ones are  given by the pair  $i,\,i+4$ mod 8,
the 16 tetrahedra split in two sets, associated to
the following cross-ratios:
\eq
t_j=(x_jx_{j-1}x_{j+1}x_{j+2})~,
~~u_j=(x_jx_{j-1}x_{j+2}x_{j-3}), ~~(j=1,2\dots8)
\label{tria16}
\en
where all the indices are taken modulo 8 and
$x_1<x_2<\dots x_8$.
Starting from the sum of the following five-term relations
written as
\begin{eqnarray}
22\;L(1)&=&L\{x_1x_2x_3x_4x_5\}+ L\{x_1x_3x_4x_5x_6\}
+\tilde L\{x_1x_2x_4x_5x_7\} + L\{x_1x_2x_3x_5x_8\}
\nn\\
&+&L\{x_5x_6x_7x_8x_1\}+ L\{x_5x_7x_8x_1x_2\}
+\tilde L\{x_5x_6x_8x_1x_3\}+L\{x_5x_6x_7x_1x_4\}
\nn
\end{eqnarray}
one can easily verify that each five-term relation
contains two tetrahedra of the triangulation
(\ref{tria16}); the remaining 24 tetrahedra combine in
pairs according to eq.~(\ref{euler1}), so one is
directly led to the sought-after  16-term relation
\eq
\sum_{i=1}^{8} \left[L(t_i)+L(u_i)\right]= 10\, L(1)
\pu
\en
Drawing the adjacency graph of these tetrahedra we
find the dual polytope, i.e. the 8-cell (or hypercube)
 represented in fig.~ 2b.
{\begin{figure}[ht]
\[\begin{array}{ll}
\hskip -1.cm
\epsfxsize=.37\linewidth\epsfbox{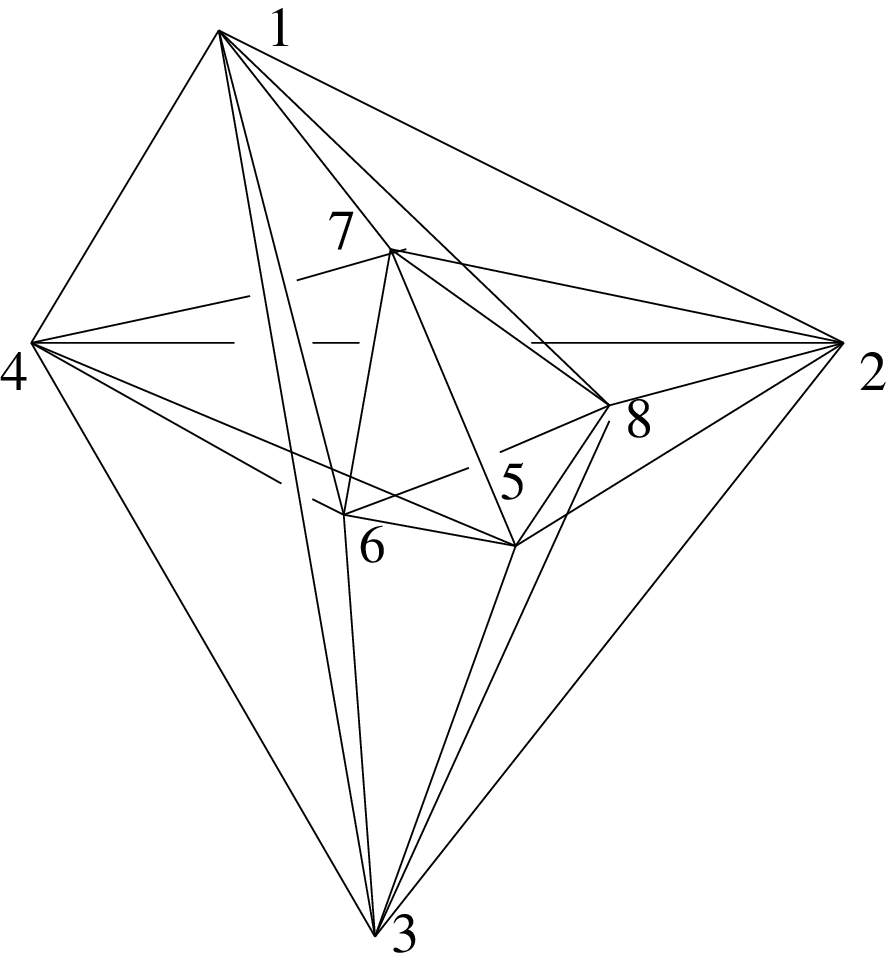}
{}&
\epsfxsize=.37\linewidth\epsfbox{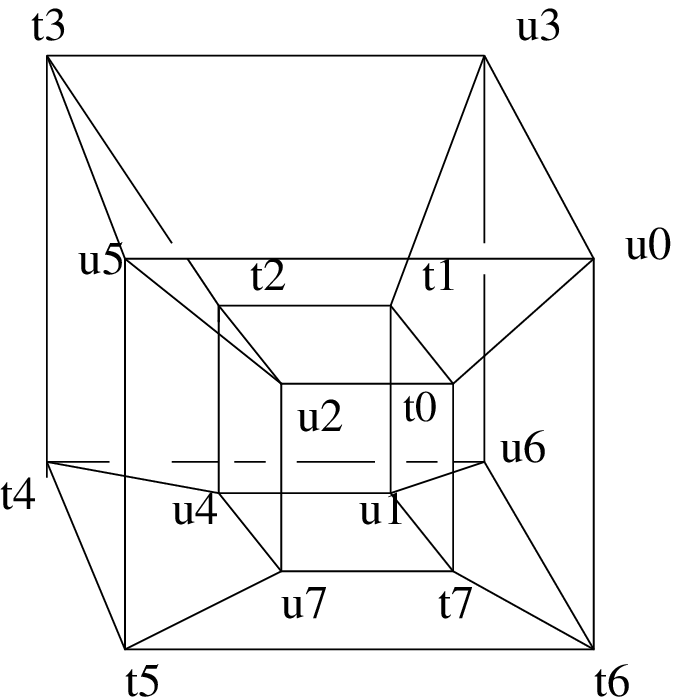}
\\
\parbox[t]{.37\linewidth}{\small 2a) $16~ cell$.}
{}~&~
\parbox[t]{.35\linewidth}{\small 2b) $8~ cell$.}
\end{array}\]
\label{cell}
\end{figure}}
There is an algebraic relation for any set of
tetrahedra connected by a closed path of such a graph.
In order to find a link with  known
Y-systems it is convenient to transform the argument
$t=(abcd)$ of $L$, using the the relation $\tilde
t=-(adcb)= t/(1-t)$. Define now
$Y_1(n)= \tilde t_{3n+3}$ and $Y_2(n)=\tilde u_{3n}$,
where all the indices are understood modulo 8.
One can  check that the following relations are
fulfilled
\begin{eqnarray}
Y_1(n-1)Y_1(n+2)&=&(1+Y_2(n))(1+Y_2(n+1))  \nn \\
Y_2(n-1)Y_2(n+1)&=&(1+Y_1(n)) / (1+1/Y_2(n))
\pu
\end{eqnarray}
This is the Y-system corresponding to the $\phi_{13}$-thermal
perturbation of  the $M_{5,8}$ minimal model.
While the discovery of the Y-system associated to the
600-cell is still an open challenge, two
infinite families of $S_3$ triangulations associated to
known integrable models may be found in ref. \cite{GT2}.
Dilogarithm  functions obey a general property known
as ``beta-map''
\cite{FS,DS,Zagier1}, that we will shortly introduce.
The beta-map and various geometrical notions relating
dilogarithms to
three-dimensional  compact manifolds will merge
together giving a more unified point of view.
Swapping between these two main tools we will then be
able to prove many of the ADE-related
dilogarithm  identities.
The beta-map condition adapted to the Rogers
dilogarithm  states that  a  sufficient condition
for having the  dilogarithm functional equations
\eq
\sum_i L(\K_i) \in  \ZZ L(1)
\virg
\label{fdilog}
\en
is that the following condition is satisfied
\eq
\sum_i
\K_i \wedge (1-\K_i)=0
\virg
\label{betamap}
\en
where the wedge product is defined through the properties
\eq
x \wedge z= -z \wedge x \virg (xy) \wedge z =
x \wedge z + y \wedge z
\label{cond}
\pu
\en
For a more  detailed  discussion of the relation
(\ref{fdilog}-\ref{betamap})
 the reader is referred to~\cite{FS,DS,Zagier1}.
If the $L$ argument  is expressed as a cross-ratio $t$,
we have $t\wedge(1-t)=(abcd)\wedge(acbd)$ and
\eq
(a b c d) \wedge (a c b d)=(a b c)-(b c d)+(c d a)-(d a b)
\virg
\label{dec}
\en
where we have defined
$$ (a b c)=(a-c) \wedge (a-b)+(a-c)
\wedge (c-b)+ (b-c) \wedge (a-b)~.$$
Comparison between eq.~(\ref{bound}) and
eq.~(\ref{dec}) shows that the beta map acts on the
cross-ratios associated to the tetrahedra like  the boundary operator
$\partial$, hence we have the implication
\eq
\partial\; (\sum_j T_j)=0 \;\Rightarrow\;
\sum_jt_j\wedge(1-t_j)=0
\virg
\label{triabm}
\en
which proves eq.~(\ref{triaid}).
It is important to note that the implication
(\ref{triabm}) cannot be inverted. Actually we found
Y-systems which do not correspond strictly
to a triangulation. The simplest example
is the  solution of the $D_N\equiv D_N \times A_1$ systems of
eq.~(\ref{YS}):
\eq
\Y_a(n+1)\Y_a(n-1)= \prod_{b=1}^N (1+ \Y_b(n))^{D^{(N)}_{ab}}
\pu
\label{YYY}
\en
These systems  emerge from
the TBA analysis of the sine-Gordon model at the
reflection-less points $\xi  =1/(N-1) $ \cite{Zf}
in the attractive regime, or equivalently
(with the  change  $\Y \rightarrow 1/\Y$)
at the points  $\xi=(N-1)$~\cite{FaZa}
in the repulsive regime.
  The solution will be written in terms of
cross-ratios of a unique
function $j\to h_j$ $(j=1,\dots N)$ exactly  as in the
models treated previously, but now $h$ is
quasi-periodic, in the sense that $h_{j-N}=q h_j+p$,
where $p$ and $q$ are real numbers.
For the $A_{N-2}$ tail of the diagram (see fig. 3a) we have
\eq
\Y_a(n)=-(h_j h_i h_{i+1} h_{j+1})~,
~~i=[\fract{n-a-2}{2}]~~,~~j=[\fract{n+a}{2}]~,~~~
a\leq N-2
\virg
\label{tail}
\en
with the square brackets indicating that the indices
must be taken modulo $N$. For the two nodes $N-1$ and
$N$ (the ``fork'' of the diagram) we can write
\eq
\Y_{N}(n \pm 1)=\Z_{\pm}(n \pm 1)~~~,~~~\Y_{N-1}(n \pm 1)=\Z_{\mp}(n \pm 1)
\virg
\label{final}
\en
with $ \Z_{+}(n )=-( h_j h_i h_{i+1} \infty ),$
$\Z_{-}(n ) = -( h_{i+1}h_{j+1} h_j \infty ),$
where $i=[\fract{n-N-1}2]$ and $j=[\fract{n+N-1}2]$.
It is now easy to verify the conjectured periodicity
$\Y_a(n+2N)= \Y_{\bar a}(n)$.
{\begin{figure}[ht]
\[\begin{array}{ll}
\hskip -1.cm
\epsfxsize=.37\linewidth\epsfbox{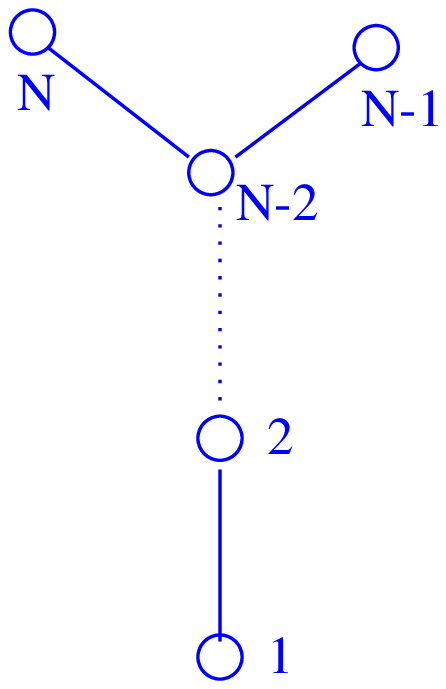}&
\epsfxsize=.47\linewidth\epsfbox{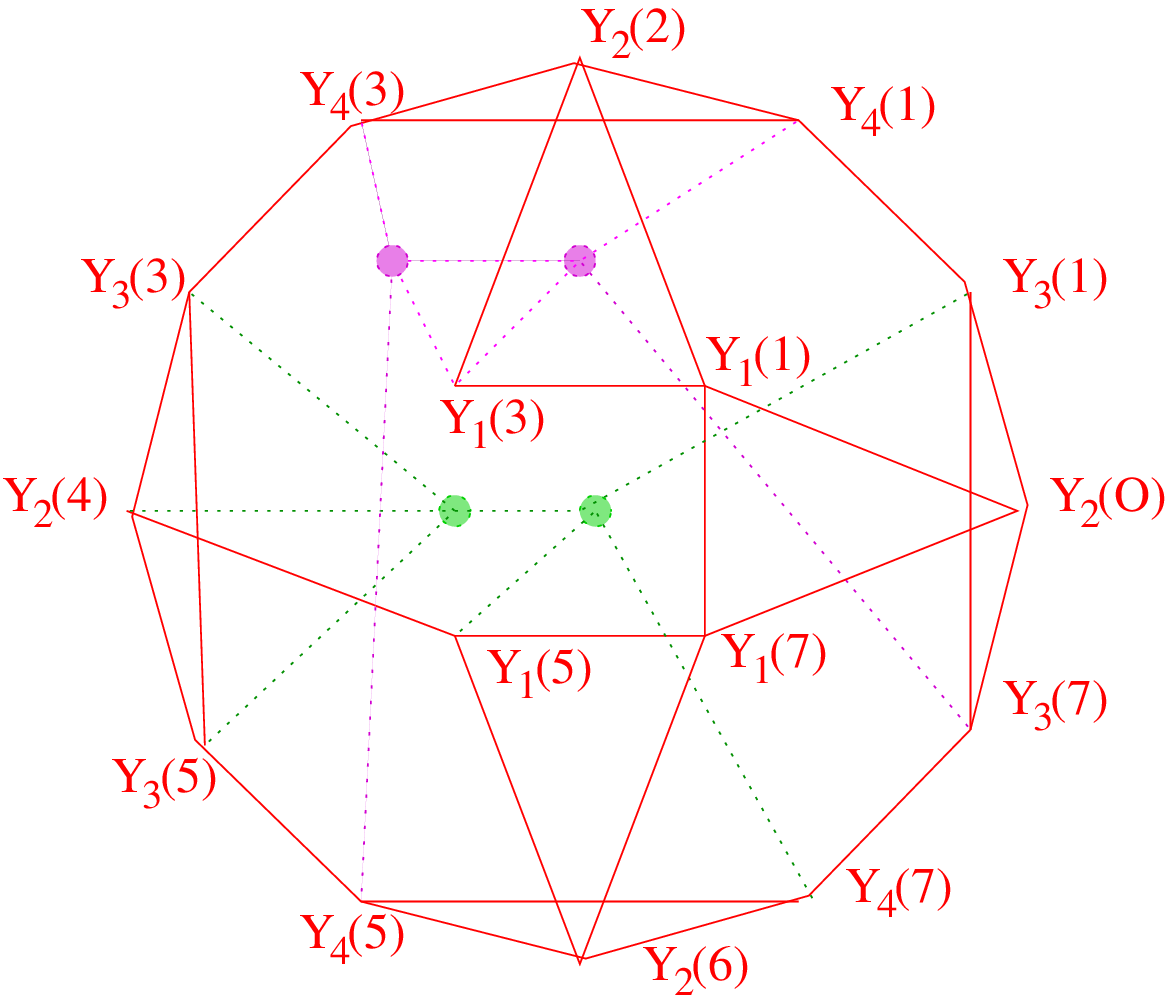}
\\
\parbox[t]{.37\linewidth}{\small 3a)~ $D_N~diagram
$.}&
\parbox[t]{.47\linewidth}{\small 3b)~ $Adjacency~
graph~ of~D_4$.}
\end{array}\]
\label{D4}
\end{figure}}
Introducing, besides the $N$ points associated to
$h_j$, $N$ further  vertices $\bar{j}=\bar 1,\dots \bar N$
associated to the values $\bar j\to qh_j+p$ and a vertex
associated to $\infty$,
it is possible to represent  this Y-system with a
3-chain $D_N$ of $N^2$ tetrahedra and $2N+1$ vertices.
Now the boundary of this chain does not vanish,
however we have $ \partial \{ D_N+B_N\}=0\;, $
where $B_N$ is the  auxiliary 3-chain
\eq
B_N=\sum_{j=1}^{N-2} \left\{[j,N,j+1,\infty] +
[\bar{j}, \bar{j+1},\bar N,\infty]\right\}
\pu
\en
We can then write a dilog  identity associated to the
triangulation $D_N+B_N$ (which turns out to be again
a triangulation of $S_3$). Since we have
$$(h_jh_Nh_{j+1}\infty)+(h_{\bar j}h_{\bar{j+1}}
h_{\bar N}\infty)=1~,$$
the tetrahedra of
$B_N$ can be paired off using the Euler identity
$L(x)+L(1-x) = L(1)$, so we end up with the result
conjectured in eq.~(\ref{adedilog}). Note that because
of the pairing just mentioned, the chain $B_N$ gives a
vanishing contribution to the beta map  (\ref{betamap}),
hence also the  chain $D_N$, although
it is not a triangulation, fulfills this condition.
As an explicit example, the adjacency graph
of the $D_4$ system is drawn in fig. 3b; the grey circles denote the
tetrahedra of  the $B_4$ chain and the dotted
lines correspond to their faces.
Other infinite families of such
``quasi-triangulations'' can be found in ref. \cite{RC}.
\section{ General ADE $\times$ ADE identities}
In this section we will  generalize the proof
of the functional dilogarithm identities to more
general cases, where the existence of an associated triangulation
is not known, using the only assumption that the associated
Y system is periodic.
Let's introduce an auxiliary set of equations  again classified in terms
of a pair of  Dynkin diagrams $(G,H)$
\eq
T_a^b(n+1) T_a^b(n-1)=
\prod_{d=1}^{r_H} T_a^d(n)^{H_{bd}}
+
\prod_{c=1}^{r_G} T_c^b(n)^{G_{ac}}
\pu
\label{Ts}
\en
When specialized to $A_N \times ADE$
the   three term relations  (\ref{Ts}) coincide
with the  well known~(cf.~\cite{KNS})
fusion-relations appearing in integrable lattice models. They are
often called    $T-systems$.
Analytically for the lower rank cases and numerically more generally
one  can again show that they fulfil the  cyclic property
\eq
T_a^b(n+P)=T_{\bar{a}}^{\bar{b}}(n)~~,~~P=h+g
\pu
\label{TP}
\en
Furthermore defining
\eq
Y_a^g(n)=\prod_{c=1}^{r_G} T_c^b(n)^{G_{ac}}
\prod_{d=1}^{r_H} T_a^d(n)^{-H_{bd}}
\virg
\label{Y_T}
\en
and
\eq
1+Y_a^b(n)= T_a^b(n+1)T_a^b(n-1)
\prod_{d=1}^{r_H} T_a^d(n)^{-H_{ba}}
\virg
\en
one can easily show  that the functions  $Y$ defined in (\ref{Y_T})
satisfy eq.s~(\ref{YS}).
Now writing the   $Y$'s in terms  of the $T$'s  using
(\ref{Y_T}) and the periodicity  (\ref{TP})  one can
show that~(\ref{betamap}) is vanishing.
We have
\begin{eqnarray}
\sum_{n=1}^P \sum_{a=1}^{r_H} \sum_{b=1}^{r_G}  Y_a^b(n)
\wedge (1+ Y_a^b(n) ) &=&
\label{ymap1} \\
\sum_{\{n,a,b\}}
( \prod_{c=1}^{r_G} T_c^b(n)^{G_{ac}} )
( \prod_{d=1}^{r_H} T_a^d(n)^{-H_{bd}} )
&\wedge&
( \prod_{m=1}^{P} T_a^b(m)^{\hat{A}^{(P)}_{mn}} )
( \prod_{e=1}^{r_H} T_a^e(n)^{-H_{be}} ) \virg \nn
\end{eqnarray}
where for convenience  the incidence matrix $\hat{A}^{(P)} $ of
an Affine $A_{P-1}$
Dynkin diagram has been introduced.
 Using $ab \wedge cd=a \wedge c + b \wedge c + a \wedge d + b \wedge d$
and
\begin{eqnarray}
\sum_{a=1}^{dim(M)} \sum_{b}^{dim(K)}
( \prod_{c=1}^{dim(M)} X_{c b}^{M_{ac}} )  \wedge
( \prod_{e=1}^{dim(K)} X_{a e}^{K_{eb}} )
&=&
 \sum_{\{a,b,c,e\}} M_{ac} K_{eb}  ( X_{cb} \wedge X_{ae})
= \nn \\
=\fract{1}{2} \sum_{\{a,b,c,e\}} M_{ac} K_{eb}
( X_{cb} &\wedge& X_{ae} -
X_{ae} \wedge X_{cb} ) = 0 \virg
\label{ymap2}
\end{eqnarray}
for $M$ and $K$ arbitrary symmetric matrices we conclude that
eq.~(\ref{ymap1}) vanishes and therefore we have
dilogarithm identities.
%
\section {Conclusions}
%
We have described a general method to write functional
dilogarithm identities associated to the TBA
formulation
of  the renormalization group evolution of integrable
quantum models.
It is worth to observe that such identities define
quantities of  topological nature, because they do
not vary under arbitrary deformations of the vertex
function $j\to x_j$, provided that the initial
ordering of the set $\{x_j\}$ is preserved.
Our geometrical description of these identities in terms
of triangulations (or quasi-triangulations) of three-dimensional
manifolds provides us with the general solution of the
recursion relations associated to the $Y$ systems and hence
a simple explanation of their periodicity. Another
consequence of this approach is to show that the $Y$'s
can be expressed in terms of a unique periodic  (or
quasi-periodic) function $x(\theta)$, related to
$x_j$ through $x_j=x(\theta+i\pi \frac jg)$, where $g$
depends on the model. In particular,
it is easy to find the $x(\theta)$ corresponding to the
stationary (i.e. $\theta$- independent) version of of
the Y systems and their associated dilog identities. For
instance one can verify, using eq.s~(\ref{tail},\ref{final})
that the stationary solution of the $A_N$ system is given by
$x(\theta)=e^{2\theta\frac{N+1}{N+3}}$ and that of the $D_N$
system is $h(\theta)=\theta $.
{}From a physical point of view, these functional identities
define a quantity which is conserved during the renormalization group
evolution of the system. In particular in the
UV limit the system is described by a conformal field theory,
then the r.h.s. of such identities must be expressible
only in terms of conformal quantities such as the
effective central charge and the
conformal dimension $\Delta$ of the perturbing operators.
For instance, in the purely elastic ADE related models
the integer $n$ of eq.~(\ref{triaid}) is given by the
ratio $\frac{c_{eff}}{\Delta}$.
The RG flow of any quantum field
theory can be viewed as a typical dissipative process,
because the integration of degrees of freedom
responsible of the trajectory between the UV and the IR
fixed points yields an information loss of the system.
Then it is amazing that there are conserved quantities
along these trajectories. It would be very interesting,
also in connection with the $c$ theorem of \ABZ,
to find a field-theoretic description of
such an invariant. \\
\noindent
{\large \bf Acknowledgements:} RT thanks   Michel Bauer  for helpful
discussions.
\renewcommand\baselinestretch{0.95}

\end{document}